\documentclass[12pt,a4paper,twoside]{article}

\newcommand{\be}{\begin{equation}}
\newcommand{\ee}{\end{equation}}
\newcommand{\bea}{\begin{eqnarray}}
\newcommand{\eea}{\end{eqnarray}}

\newcommand{\Ima}{{\rm Im}}

\begin{document}

\begin{titlepage}
\begin{flushleft}
       \hfill                      {\tt hep-th/9905227}\\
       \hfill                      UUITP-10/99\\
       \hfill                       May 1999\\
\end{flushleft}
\vspace*{3mm}
\begin{center}
{\bf {\Large Spherically Collapsing Matter in AdS, \\
Holography, and Shellons\\}}
\vspace*{12mm}
{\large
Ulf H. Danielsson\footnote{E-mail: ulf@teorfys.uu.se} \\
Esko Keski-Vakkuri\footnote{E-mail: esko@teorfys.uu.se}\\
Mart\'{\i}n Kruczenski\footnote{E-mail: martink@teorfys.uu.se} \\

\vspace{5mm}

{\em Institutionen f\"{o}r teoretisk fysik \\
Box 803\\
S-751 08  Uppsala \\
Sweden \/}}
\vspace*{15mm}
\end{center}

\begin{abstract}
We investigate the collapse of
a spherical shell of matter in an anti-de Sitter space. We search for a holographic
description of the collapsing shell in the boundary theory. It turns
out that in the boundary theory it is possible to find information
about the radial size of the shell. The shell deforms
the background spacetime, and the deformed background metric enters
into the action of a generic bulk field. As a consequence, the
correlators of operators coupling to the bulk field are modified.
 By studying the analytic structure
of the correlators, we find that in the boundary theory there are
unstable excitations ("shellons") whose masses are multiples  of a scale set
by
the radius of the shell. We also comment on the relation between
black hole formation in the bulk and thermalization in the boundary.

\end{abstract}

\end{titlepage}

\baselineskip 16pt

\section{Introduction}

One of the many \cite{Aharony:1999ti} exciting aspects of the AdS/CFT
correspondence \cite{Maldacena:1997re, Gubser:1998bc, Witten:1998qj} is that
it contains the seeds for solving the black hole information puzzle \cite
{Hawking:1976ra} in a holographic \cite{'tHooft:1993gx, Susskind:1995vu}
way. So far, the understanding of black holes in the context of the AdS/CFT
correspondence is mostly limited to the properties of \emph{static} solutions.
For example, one can give a boundary theory interpretation for the
Bekenstein-Hawking entropy \cite{Strominger:1998eq, Birmingham:1998jt},
properties of different vacua and their thermodynamic interpretation \cite
{Maldacena:1998bw, Balasub:1998sn, Balasub:1998de, Keski-Vakkuri:1998nw,
Danielsson:1998wt, Spradlin:1999bn}. However, the static solutions have some
rather special features, such as the appearance of multiple asymptotic
regions\footnote{However, see \cite{Louko:1998hc}.}. To address the issue
of the fate of information, one would like to
study black holes which form from collapsing or colliding matter\footnote{%
Black hole creation in AdS$_{3}$ has been studied in \cite{Peleg:1995wx,
Matschull:1998rv, Holst:1999tc}. In these cases the resulting black holes
are stable.}, and their subsequent evolution to the end. These issues are
much less clear as discussed in \cite{Polchinski:1999yd}.
In this paper we take a first step towards understanding the collapse
of a spherical shell of matter in AdS space.

The first problem to confront is how to account for the presence of a shell
of matter in the bulk of an AdS space. In particular, how does one obtain information
about the radial size of a shell? For a pointlike probe, the scale-radius
duality \cite{Susskind:1998dq, Banks:1998dd, Balasub:1998de,
Peet:1998wn}
converts its radial distance in the bulk into a transverse scale
of its image at the boundary. However, in the case of a shell, the
radial scale is a bit trickier to find because of spherical symmetry.
What is the holographic image of a shell at the
boundary? Since a pure AdS space corresponds to a vacuum state in the
boundary, and the end result of the shell collapse (if it is a stable black
hole with positive specific heat) corresponds to a boundary theory at finite
temperature \cite{Witten:1998qj, Hawking:1983dh}, a first guess is that
adding a shell to the AdS space corresponds to an off-equilibrium
configuration in the boundary theory, which then evolves towards thermal
equilibrium as the shell collapses to a black hole in the bulk (see e.g.
\cite{Balasub:1998de, Horowitz:1999gf} for discussion). Further, the
scale-radius duality of the AdS/CFT would suggest that the initial
configuration is somehow peaked at higher energies than the temperature of
the black hole. In other words, one might hope to associate an initial
average occupation number spectrum with the initial configuration, with a
peak at energies corresponding to the initial radius of the shell, and study
how the time evolution rearranges the spectrum to a thermal one, with the
peak moving down in energies to stabilize at the Hawking temperature.

To test the above idea, one would need some way of detecting or computing
e.g. the average occupation number spectrum in the boundary theory. One
would then need to study operators which have access to the excited states.
For example, the shell deforms the background spacetime, and the deformation
couples to the stress-energy tensor of the boundary theory, giving it an
expectation value. The expectation value gives information only about the
total energy, not about the details of the distribution. However, one could
study higher n-point correlators of the stress tensor; these involve more
detailed information about the spectrum which one can then try to extract
out\footnote{In general, higher n-point correlation
functions are needed to have more complete information about
a quantum state. For example, in \cite{Polchinski:1999yd} it was
discussed how higher n-point functions are relevant for understanding
the causal structure in AdS/CFT.}.

Another, more direct, way to detect the shell would be to assemble it from
matter associated with a bulk field coupling to an operator $\mathcal{O}_{%
\mathrm{shell}}$ in the boundary. Then, the n-point functions of
$\mathcal{O}%
_{\mathrm{shell}}$ would provide information of the shell itself, without a
detour to its backreaction to the metric. However, this route obviously
depends on what the shell is made of and how.

In this paper, we shall follow a third possible strategy. Since the shell
deforms the background spacetime, the action for any bulk field in the
deformed background is modified as well. Since the bulk action is the
generating function for the correlation functions of the corresponding
operators in the boundary theory, the correlation functions will reflect the
presence of the shell. Even if the expectation values of the operators (the
one point functions) vanish, the 2-point functions can be different from
those evaluated in the vacuum. For example, in the end of the collapse, when
the boundary theory thermalizes, a two-point function will become periodic
in imaginary time, even if the operator has a zero expectation value in the
thermal background. So our strategy is that we shall not try to directly
probe the states associated to the shell, but we investigate their effect on
other operators in the boundary theory, and use their correlation functions
to obtain information about the shell. A related possibility would be to
study the properties of a hanging string
\cite{Rey:1998ik, Maldacena:1998im, Brandhuber:1998er, Rey:1998bq}.
In the case of a black
hole the boundary theory will be at finite temperature, and the force
between two quarks will be screened at sufficiently large separation. In the
picture with the quarks corresponding to endpoints of a hanging string,
screening occurs when the string breaks and one obtains a configuration with
a string from each of the quarks going straight down through the horizon of
the black hole. With a shell the situation is different since the boundary
state is not thermal and screening is not expected. From the bulk point of
view there is no horizon and the string between the quarks can therefore not
break. Even though the distance dependence of the force will be affected by
the presence of the shell, there is still a long range contribution.

The simplest two-point function is given by a bulk scalar field $\phi $
coupling to an
operator $\mathcal{O}$ in the boundary theory. For example, $\phi $ could be
the five dimensional dilaton in AdS$_{5}$, and $\mathcal{O}$ the operator $%
\mathrm{Tr}F^{2}$ in the 3+1 dimensional $N=4$ SYM theory. Also, we will
concentrate on the simplest possible situation. We consider a spherically
symmetric thin shell of matter, and focus on the initial stage of the
collapse when the shell is moving very slowly. In that case, we can use a
quasistatic approximation and take the shell radius to be fixed. We solve
the field equation of $\phi $ on both sides of the shell, impose appropriate
boundary conditions, and join the solutions across the shell. We will then
derive the two point function\footnote{%
Note that in the case of the static shell, the two-point function will
depend only on the differences $(t-t^{\prime })$ and $(\vec{x}-\vec{x}%
^{\prime })$.}
\[
\langle \mathcal{O}(t,\vec{x})\mathcal{O}(t^{\prime },\vec{x}^{\prime
})\rangle
\]
for the corresponding boundary operator $\mathcal{O}$. Actually, our
calculation will give the momentum space Fourier transform of the two-point
function,
\begin{equation}
G(\omega ,\vec{k})=\int dt\int d^{d}\vec{x}\ e^{-i\omega t+i\vec{k}\cdot
\vec{x}}\langle \mathcal{O}(t,\vec{x})\mathcal{O}(0,0)\rangle \ .
\end{equation}

The analytic structure of the propagator gives information on the spectrum
of excitations of the theory. In the presence of the shell, the two point
function will be identified as the retarded propagator $G_{\mathrm{ret}}
(\omega ,\vec{k} )$. For a fixed value of $\vec{k}$, the retarded propagator
is analytic in the upper complex $\omega$ plane. It can be written in an
integral representation form
\begin{equation}
G_{\mathrm{ret}} (\omega, \vec{k} ) = \int^{\infty}_{-\infty} \frac{%
d\omega^{\prime}}{2\pi} \frac{A(\omega^{\prime},\vec{k})}{\omega
-\omega^{\prime}+i\epsilon} \ ,  \label{intrep}
\end{equation}
where $A(\omega ,\vec{k} )$ is the spectral density function. It is the
imaginary part of the retarded propagator,
\begin{equation}
A(\omega ,\vec{k} ) = -2 \cdot \mathrm{Im} G_{\mathrm{ret}} (\omega ,
\vec{k}
) \ .  \label{spectral}
\end{equation}
(See \emph{e.g.} \cite{mahan} for discussion.) The spectral density function
can be interpreted as the probability\footnote{%
For bosonic excitations, one needs $\omega>0$.} that an excitation created
by the operator $\mathcal{O}$ has an energy $\omega$ and a momentum
$\vec{k}$. Let us now recall some possible
analytic features of the retarded propagator in a generic QFT.
If we try to do an analytic continuation to the lower half-plane, and take
the limit $\epsilon \rightarrow 0$, we may find that the propagator has
poles at the real $\omega$ axis. In the boundary theory, these correspond to
elementary excitations or stable bound states (the latter being possible in
an interacting theory). The poles of the propagator correspond to delta
function peaks of the spectral density function. There may also be cuts,
signaling multiparticle states\footnote{Of course, this issue played
an important part in studies of absorption by branes \cite{Gubser:1997yh},
in the path which led to the AdS/CFT correspondence. See
\cite{Aharony:1999ti}
for more discussion and references.}. We may also find poles of the
propagator
in
the lower complex plane, corresponding to unstable bound states
(resonances). If we denote the location of such a pole by $\omega_i(\vec{k}%
)-i\Gamma_i$, the imaginary value $\Gamma_i$ is the width of the unstable
bound state. The closer to the real axis the pole is, the longer lived the
resonance is. In the spectral density function, the resonances appear
smeared peaks with finite width. (See \emph{e.g.} \cite{pines,mahan} for
discussion).

Thus, the analytic structure of the propagator gives information on the
stable and unstable excitations present in the boundary theory. In general,
the operators $\mathcal{O}$ are composite operators built out of the
elementary fields of the boundary theory. Hence, we obtain only information
about a certain subset of particles propagating in the boundary, the stable
and unstable composite objects that can be created and annihilated by the
operator $\mathcal{O}$. 

The key fact is that this gives us enough information to detect
the radial size of the shell.
We discover unstable resonances in the boundary theory, whose
masses are multiples of a mass scale set by the radial size of the 
shell. The boundary theory is conformal, but it is in
an off-equilibrium state characterized by a dimensional parameter, so the
massive resonances can exist.

The resonances are discussed in Section 2. In section 3 we discuss the results.

\section{Propagators in the shell background}

We shall now proceed, step by step, to compute the two point function of a
boundary operator in the case that a (quasi)static shell of matter is
present in the bulk. Although this is an unphysical situation, it can be
used as an approximate description of a slowly moving shell. In \cite
{Witten:1998qj, Gubser:1998bc}, it was described how to compute boundary
correlators of an operator $\mathcal{O}(x,t)$ using classical solutions for
a related bulk field $\phi(r,x,t)$ using the AdS/CFT correspondence. The
computation method we will use is a variant of the one described in the
appendix of \cite{Freedman:1998tz}.

In all cases, the spacetime is asymptotically $AdS_{d+1}$ ($d=2,4$), and
near the boundary the bulk metric reduces to the approximate form
\begin{equation}
ds^2 \approx _{r\rightarrow\infty} -\frac{r^2}{R^2} dt^2 + \frac{R^2}{r^2}
dr^2 + \frac{r^2}{R^2} d\vec{x}^2  \label{poin}
\end{equation}
where the boundary is at $r\rightarrow\infty$. Let $\phi$ be a scalar field
in the bulk, with a mass $m$. In Minkowski signature, the classical equation
of motion
\begin{equation}
\frac{1}{\sqrt{g}} \partial_{\mu}\left(\sqrt{g}g^{\mu\nu}\partial_{\nu}
\phi(r,\vec{x},t)\right) - m^2 \phi(r,\vec{x},t)=0  \label{eom}
\end{equation}
has two kinds of solutions: normalizable and non-normalizable ones. We
denote the former by $\phi^{(+)}(r,\vec{x},t)$ and the latter by $%
\phi^{(-)}(r,\vec{x},t)$; they have an asymptotic behavior
\begin{equation}
\phi^{(\pm)}(r,\vec{x},t) \rightarrow r^{-2h_{\pm}} \ \ \ \ (r\rightarrow
\infty )
\end{equation}
where
\begin{equation}
h_{\pm} = \frac{1}{4}(d\pm \sqrt{d^2 + 4m^2} ) \equiv \frac{d}{4}\pm \frac{%
\nu}{2} \ .
\end{equation}
The starting point is to impose appropriate boundary conditions for the bulk
field $\phi$ at the interior of the AdS space. In the asymptotic region, the
field will in general become a superposition of the normalizable and
non-normalizable solutions, with an asymptotic behavior
\begin{equation}
\phi(r,\vec{x},t) \approx r^{-2h_+} \phi_+ (t,\vec{x} ) + r^{-2h_-} \phi_-
(t,\vec{x} ) \ .  \label{asy}
\end{equation}
In the above, $\phi_-(t,x)$ is the boundary data which acts as a source for
an operator $\mathcal{O}(t,x)$ in the boundary theory. Next, we Fourier
transform the variables $t,x$ and write the equation (\ref{asy}) in the form
\begin{equation}
\phi(r,\omega,\vec{k} ) \approx \left( r^{-2h_-} + r^{-2h_+} G(\omega ,
\vec{%
k} ) \right) \phi_- (\omega, \vec{k} ) \ ,
\end{equation}
where
\begin{equation}
G(\omega, \vec{k} ) = \frac{\phi_+(\omega , \vec{k} )}{\phi_-(\omega ,
\vec{k%
} )} \ .  \label{gee}
\end{equation}
One can check that the two point function in the boundary will be
\begin{equation}
\langle \mathcal{O} (\omega , \vec{k} ) \mathcal{O} (\omega^{\prime},
\vec{k}%
^{\prime}) \rangle =(h_+-h_-) \delta(\omega+\omega^{\prime}) \delta
(k+\vec{k%
}^{\prime}) G(\omega , \vec{k} ) \ .
\end{equation}
Thus, the essential steps of the computation are as follows: one imposes
interior boundary conditions for the bulk field, writes it as a
superposition of the normalizable and non-normalizable modes, then the ratio
of the superposition coefficients gives the two point function in the
boundary. To obtain the answer with a correct overall coefficient
$(h_+-h_-)$%
, one needs to perform the calculations with more subtlety, as discussed in
\cite{Freedman:1998tz}. However, overall coefficients will not be important
here, so they will be suppressed in what follows. The field $\phi(r,\vec{x}%
,t)$ must also obey time boundary conditions related to the fact that one
can compute, for example, retarded, advanced or Feynmann propagators. In the
absence of interactions, and with respect to a vacuum, all propagators in
momentum space are equal. Otherwise, the propagators can be identified
by examining their analytic structure.

As a warm-up, we consider the propagator in the absence of a shell in the
bulk. The full AdS$_{d+1}$ manifold is covered by global coordinates, with
the metric
\begin{equation}
ds^2 = -\frac{r^2 + R^2}{R^2}dt^2 + \frac{R^2}{r^2 + R^2} dr^2 +
\frac{r^2}{%
R^2} d\Omega^2_{d-1}  \label{global}
\end{equation}
where the radial coordinate $r$ extends from $0$ to $\infty$
and $d\Omega^2_{d-1}$ is the metric of a $(d-1)$-sphere of radius $R$. 
The interior
boundary condition is that $\phi$ must be regular at the origin $r=0$. The
solution is expressed as a linear superposition of normalizable and and
nonnormalizable modes in \cite{Balasub:1998sn}. The propagator (\ref{gee})
is simply the ratio of the relative coefficients, which are found in
equation (37) of Ref. \cite{Balasub:1998sn}. The answer is
\begin{equation}
G(\omega ,k ) = \frac{\Gamma(-\nu)}{\Gamma(\nu)} \frac{
\Gamma(h_++\frac{R}{2}(\omega+k))\Gamma(h_++\frac{R}{2}(-\omega+k))}
{\Gamma(h_-+\frac{R}{2}(\omega+k))\Gamma(h_-+\frac{R}{2}(-\omega+k))} \ ,
\label{gloprop}
\end{equation}
where $k={\rm integer}/R$.
The above result is valid when $\nu$ is not an integer. If $\nu$ is an
integer then the propagator can be written in terms of derivatives of
$\Gamma
$ functions. In this case it is easier to compute first the spectral
function part and then use the integral representation (\ref{intrep}) to
reconstruct the full propagator. It is easy to see that the propagator (\ref
{gloprop}) has poles at $\omega \pm k= \mp(2n+h_+)\frac{1}{R}$,
$n=1,2,\ldots$ corresponding to the normalizable modes in AdS space. 
From the boundary point of view the discrete spectrum 
is possible since the spatial directions are 
compactified to a sphere $S^{d-2}$.  Accordingly, note that the
dispersion relation is not Lorentz invariant.  Between successive
poles, the propagator (\ref{gloprop}) has also zeroes at $\omega \pm k=
\mp(2n+h_-)\frac{1}{R}$.

In the limit $R\rightarrow \infty$, keeping $\omega$ and $k$ fixed, the
poles and zeros accumulate and they degenerate to cuts in
the
propagator for $\omega>k$ and
$\omega<-k$. Further, in the limit $R\rightarrow \infty$ the boundary
decompactifies. Applying the limit to (\ref{gloprop}%
),  we find that the propagator reduces to the form
\begin{equation}
\ G(\omega, k) \sim (\omega^2-k^2)^\nu \ \ \ \ (\omega^2 \gg k^2 ) \ .
\label{poinprop}
\end{equation}
Thus, we recover the expected result for a free propagator in an infinite
boundary. From the bulk point of view one has to take $R\rightarrow \infty$ with 
$r/R^2$ fixed or equivalently $r\rightarrow \infty$. In this limit the metric 
reduces to (\ref{poin}). Using this metric the propagator (\ref{poinprop})
follows immediately. The only subtlety is the boundary condition to use 
in the interior since in the large $R$ limit, the point $r=0$
corresponds to the coordinate horizon of a Poincare patch. 
Near the horizon any solution can be written as a linear combination
of an ingoing and an outgoing wave. If $\omega$ has a small imaginary part
it is easily seen (see Appendix) that one has to use an ingoing wave 
for $\Ima(\omega)>0$ and an outgoing one for $\Ima(\omega)<0$.
The use of different boundary conditions for $\Ima(\omega) >0$ and $\Ima(\omega)<0$ 
produces a cut on the real axis corresponding to switch from the retarded
to the advanced one.

Let us now turn to the case of a black hole of radius $r_+>R$
in AdS space. Now we must impose
a boundary condition at the horizon. Again, both solutions of the wave
equation are regular at the horizon. The general solution is a superposition
of an outgoing and an ingoing wave:
\be
 \phi \approx A (r-r_+)^{{\rm i}\frac{\omega}{2}}
      + B (r-r_+)^{-{\rm i}\frac{\omega}{2}} \ \ \  (r\rightarrow r_+).
\ee
Imposing the condition that $\phi$ be regular at $r=r_+$ implies
that if $\Ima \omega>0$ we take the solution with $A=0$ and
with $B=0$ if $\Ima \omega<0$. Since
the solutions with $A=0$ and with $B=0$ are one complex conjugate
of the other, the imaginary part of the propagator changes sign
when $\omega$ crosses the real axis.
In the case of a BTZ black hole of radius $r_{+}$, the propagator
(\ref{gee})
follows from\footnote{For coordinate space expressions including the
correct overall coefficient, see also \cite{Muller-Kirsten:1998mt,
Ohta:1998xh}.}
\cite{Keski-Vakkuri:1998nw, Ichinose:1995rg, Teo:1998dw, Lee:1998bf,
Birmingham:1997rj, Maldacena:1997ix}
and is
\begin{equation}
G_{\mbox{ret}}(\omega ,k)=\lim_{\epsilon\rightarrow 0}
G(\omega+{\rm i}\epsilon ,k)=\frac{\Gamma (1-\nu )}{\Gamma (1+\nu )}
\frac{
 \Gamma (h_+-\frac{{\rm i}R}{2}(\omega+k))
 \Gamma (h_+-\frac{{\rm i}R}{2}(\omega-k))}
{\Gamma (h_--\frac{{\rm i}R}{2}(\omega+k))
 \Gamma (h_--\frac{{\rm i}R}{2}(\omega-k))}  \label{BTZprop}
\end{equation}
As before, if $\nu $ is integer then the calculation 
must be done more carefully. 
For example, for $\nu =1$ the result is
\begin{equation}
G(\omega ,k)=\frac{(\omega ^{2}-k^{2})}{T^{2}}\frac{1}{e^{\frac{\omega
+k}{2T%
}}-1}\frac{1}{e^{\frac{\omega -k}{2T}}-1}
\end{equation}
This propagator can be understood as a thermal propagator of a composite
operator. Acting on a vacuum, the operator $\mathcal{O}$ creates states with
right- and leftmoving particles of two-momentum $k_{1}=(\omega _{1},\omega
_{1})$ and $k_{2}=(\omega _{2},-\omega _{2})$. These two-particle states
have total energy $\omega =\omega _{1}+\omega _{2}$ and total momentum $%
k=\omega _{1}-\omega _{2}$. With this notation the propagator reads
\begin{equation}
G(\omega ,k)=\frac{1}{T^{2}}\frac{\omega _{1}}{e^{\frac{\omega _{1}}{T}}-1}%
\frac{\omega _{2}}{e^{\frac{\omega _{2}}{T}}-1}\ .
\end{equation}
Now it is obvious that this is a thermal propagator for a composite operator
propagating two particle states. The Bose occupation numbers arise from \
the gamma functions \cite{Maldacena:1997ix}. In fact this propagator follows
simply from conformal invariance, so it does not give us any information
about interactions in the boundary theory, only that it is at temperature $%
T\sim r_{+}/R^2$. A similar calculation can be done for the black hole in $%
AdS_{5}$ using the function obtained in \cite{Danielsson:1998wt}. The result
is similar to the case of $AdS_{3}$ although it cannot be expressed in terms
of special functions as in (\ref{BTZprop}).

After this lengthy introduction, we are prepared to compute the propagator
in the presence of a shell of radius $r_{s}$. In the interior of the shell,
the metric is the AdS metric in global coordinates, given by (\ref{global}).
The exterior metric is the AdS black hole metric,
\begin{equation}
ds^{2}=-(1-\frac{\mu }{r^{d-2}}+\frac{r^{2}}{R^{2}})dt^{2}+\frac{dr^{2}}{(1-
\frac{\mu }{r^{d-2}}+\frac{r^{2}}{R^{2}})}+\frac{r^{2}}{R^{2}}d\Omega
_{d-1}^{2}\ ,  \label{adsbh}
\end{equation}
where the parameter $\mu $ is related to the ADM mass of the black hole and
determines the horizon radius $r_+$. The
interior boundary condition for the field $\phi $ is the same as in pure AdS
space: regularity at $r=0$. 
The equation (\ref{eom}) reduces to:
\begin{equation}
\frac{1}{r^{d-1}}\partial _{r}\left( r^{d-1}f(r)\partial _{r}\phi (r,\omega
,k)\right)
+\left( \frac{\omega ^{2}}{f(r)}-\frac{k^{2}}{r^{2}}-m^2\right) \phi
(r,\omega
,k)=0  \label{radeq}
\end{equation}
with
\begin{equation}
f(r)=\left\{
\begin{array}{lcl}
f_{1}(r)=1+r^{2} & \mbox{if} & r<r_{s} \\
f_{2}(r)=1-\frac{\mu }{r^{d-2}}+\frac{r^{2}}{R^2} & \mbox{if} & r>r_{s}
\end{array}
\right.
\end{equation}
and $\frac{k^2}{R^2} = l (l+d-2), l\in Z_{\ge 0}$ is the eigenvalue of the
laplacian
on the sphere $S^{d-1}$.
The solutions of the field equation, $\phi _{1}$ for $r<r_{s}$ and $\phi
_{2}
$ for $r>r_{s}$, are the known solutions in the global and black hole
coordinates. The matching conditions at $r=r_{s}$ are
\begin{eqnarray}
\left. \phi _{1}\right| _{r=r_{s}} &=&\left. \phi _{2}\right| _{r=r_{s}} \\
\left. f_{1}(r)\partial _{r}\phi _{1}\right| _{r=r_{s}} &=&\left.
f_{2}(r)\partial _{r}\phi _{2}\right| _{r=r_{s}}
\end{eqnarray}
where the last condition follows from integrating the equation between $%
r_{s}-\epsilon $ and $r_{s}+\epsilon $ with $\epsilon \rightarrow 0$. It
also ensures current conservation. Using the matching conditions the
propagator follows as
\begin{equation}
G(\omega ,k)=-\frac{f_{2}\phi _{1}\partial _{r}\phi _{2}^{(-)}-f_{1}\partial
_{r}\phi _{1}\phi _{2}^{(-)}}{f_{2}\phi _{1}\partial _{r}\phi
_{2}^{(+)}-f_{1}\partial _{r}\phi _{1}\phi _{2}^{(+)}}  \label{match}
\end{equation}
where $\phi _{2}^{(\pm )}$ are the normalizable and non-normalizable modes
in the black hole background (in three dimensions, they can be found in
\cite
{Keski-Vakkuri:1998nw}), and $\phi _{1}$ is the mode in global coordinates
that is regular at $r=0$. We have performed this calculation, and have found
a tower of resonances in the boundary theory. However, it will be much
simpler and more illustrative to discuss a computation which uses some
additional approximation methods, which are good for a large shell with
radius $r_{s}\gg r_{+} \gg R$. In this limit, the metric in the interior
reduces to the Poincare metric, so the solution in the interior will be a
Poincare mode. Correspondingly, the interior boundary condition will be
mapped to those in a Poincare patch, as we discussed before in the context
of the large $R$ limit. We will relegate the details in the Appendix, and
quote just the end result. By studying the propagator in $d$ dimensions, we
find that it has an infinite number of poles at the complex values
\begin{equation}
\omega _{n}\approx \frac{\pi
r_{s}}{R^2}\left( n+\frac{3}{4}+\frac{\nu }{2}\right) 
 -\frac{ir_{s}}{2R^2}
 \ln \left( \frac{4\pi n}{(d-1)}\frac{r_{s}^{d}}{r_{+}^{d}}\right) \ .
\end{equation}
As we discussed in the Introduction, these poles are interpreted as unstable
excitations (resonances) in the boundary theory that appear due to the shell
background. The real part (the mass) is proportional to the radius of the
shell. We call the unstable excitations as ''shellons''. Note also that the
width is also of the order of the shell radius, hence the shellons have a
short lifetime. This is good, since then a slowly moving shell is static
compared to the lifetime of the shellons, and the results are consistent
with the quasistatic approximation.

\section{Discussion}

\bigskip

Our calculation shows the existence of unstable excitations, shellons, with
properties like mass and life time depending on the radius of the shell. 
Thus, in particular, one can detect a difference between spherical shells
of different radii but with the same mass.
It would be very interesting to have a more detailed understanding of how the
shellons and their properties are to be understood from the boundary point
of view.

\bigskip

Another important problem is to let the shell fall freely and follow its
evolution into a black hole. Let us make some comments on what to expect,
but first we need a more detailed understanding of the composition of the
shell from the boundary point of view. We begin with a static shell
consisting of a large number of elementary particles. A single particle \
with mass of the order $1/R$ at rest at radius $r_{s}$, will through
holography show up as a blob in the boundary with size $R^{2}/r_{s}$, and
total energy  given by $E\sim r_{s}/R^{2}$. The shell will therefore be \ a
superposition of such blobs, and correspond to a state that is out of
thermal equilibrium with a homogeneous energy density. If we turn to the
falling shell, each blob will turn into a bubble expanding at the speed of light
\cite{Danielsson:1998wt}.
and a
falling dust shell containing several particles will then correspond to a
superposition of such expanding bubbles. How will they interact?

To understand this, it is useful to consider first a particle falling into a
black hole.
In \cite{Danielsson:1998wt} the shape of a bubble produced by a particle
falling into a BTZ black hole was given. If one considers the energy
momentum
tensor
as in \cite{Horowitz:1999gf} then the shape is given by\footnote{This
follows
simply
by performing the conformal transformation that corresponds to change from
global
to BTZ coordinates in the bulk} :
\be
T_{00} =
\frac{1}{\left(a^2+(1+a^2)\sinh^2\frac{r_+(x+t)}{2}\right)^{2}}+
\frac{1}{\left(a^2+(1+a^2)\sinh^2\frac{r_+(x-t)}{2}\right)^{2}}
\label{eq:BTZbubble}
\ee
where $a$ is a parameter related to the initial position of the particle and
$r_+$ is the radius of the black hole.
We see then that in this case the bubble still propagates at the speed of light.
However, for fixed $t$, the energy density decays exponentially for large $x$ with a 
characteristic length of $1/T$ where $T\sim r_+/R^2$ is the temperature of the black hole,
while if no thermal bath is present, it decays with a power law.
If the particles
consist of supergravity modes, their mass is of order $\sim 1/R$ while that
of a
black hole
is $\sim N^2/R$ and so we need order $N^2$
of them. In the large $N$ limit the gas of bubbles in the boundary will be
very
dense
and we expect that each bubble will see a mean field created by the rest.
Each
bubble
will then expand similarly as in eq.(\ref{eq:BTZbubble}). Alternatively, we
can
have a
dilute gas of bubbles in the boundary
if the
shell consists of $k$ dust particles, each containing of the order $N^{2}$
elementary particles of mass $1/R$. This could be thought of as a shell
consisting of $k$ black holes (of course in this case it is not a thin
shell).
In the boundary, one now finds a dilute
gas of $k$ expanding bubbles.  Even when the falling dust approaches
$r_{+}$%
, it is easy to see that the bubbles still remain dilute, unless the
boundary equivalent of the bulk gravitational interactions are taken into
account. The interactions will, as the final black hole forms, merge all the
bubbles into a homogeneous background. Another way to understand this is to
note that the holographic scale/radius relation is changed when the
deviation from a pure AdS metric becomes important, making the bubbles even
larger. Does this mean that the $k$ small black holes have merged into one
big black hole? In this context it is important to know the strength of the
interactions. The interactions between the black holes that make up the
shell are not suppressed by powers of $N$, and are therefore part of the
supergravity description. In units where $R=1$, the Newton's constant is $%
G\sim 1/N^{2}$, but since the mass of each of the black holes that make up
the shell is $M\sim N^{2}$, \cite{Horowitz:1999gf} the corrections to the
metric
around
one of the black holes are $GM\sim 1$. However, we do not expect these
collective interactions to result in a final thermal state, since this will
require interactions between the individual particles that make up the black
holes and the shell. An interaction between any two elementary particles
will be suppressed by powers of $N$, and therefore thermalization is
expected to be a $1/N$ effect. We expect this to be a general result, not
depending on the detailed composition of the collapsing shell. It is
intriguing that similar results have been  obtained previously in studies of
thermalization in large $N$ field theories in a different context
\cite{Boyanovsky:1995me, Cooper:1997ii, Boyanovsky:1998ba}. In fact
the collapse of the falling shell gives interesting information
about the non-equilibrium processes they have studied. However a precise
relation is certainly lacking.

\bigskip

While it is satisfying that we can find information in the boundary theory
about a collapsing shell in the bulk, it is also obvious that this is just a
step towards understanding the formation of black holes as a hologram.
One
of the steps to follow would be to go beyond the quasistatic approximation
and study what happens as the shell is falling rapidly, or what happens when
it crosses its Schwarzschild radius.
It will also be interesting to study
what happens when the resulting black hole is small and has a negative
specific heat. Another interesting issue would be to add charge to the
collapsing matter, to make contact with the recently discussed charged AdS
black hole solutions \cite{Chamblin:1999tk, Chamblin:1999hg}.

\bigskip\

\bigskip

\section*{Acknowledgements}

We would like to thank Jorma Louko for many useful discussions. 
We also thank Per Kraus for helpful comments on a draft of this paper.

\section*{Appendix}

We are interested in finding the solutions in the region near the radius
of the shell, where we match the interior solution with the exterior
solutions.
If the shell is large, this region is at $r\gg R$, where the interior metric
takes the approximate form (\ref{poin}). Correspondingly, the interior
solutions
are equal to the Poincare modes, which can be expressed using Bessel
functions,
as discussed {\em e.g.} in \cite{Balasub:1998sn}.
The situation is now analogous to the large $R$ limit which we discussed in
section 2. The continuum of Poincare modes will correspond to a cut in the
propagator. Correspondingly, we have a choice of two interior boundary
conditions:
the solution must look like an ingoing or outgoing wave as $r\rightarrow 0$.
Which one is appropriate follows by taking the unique solution in global
coordinates
and obtaining its behaviour for large frequencies. Alternatively one can
reason
as follows:
an ingoing wave behaves as
\be
\phi_1(r) \approx r^{\frac{1-d}{2}} e^{{\rm
i}\frac{\sqrt{\omega^2-k^2}}{r}},\
\ (r\rightarrow 0)
\ee
where we set $R=1$ as in the rest of the Appendix.
If $\omega$ has a small positive imaginary part the function is well behaved
for
$r\rightarrow 0$ but it blows up if the imaginary part is negative. So
an ingoing wave is appropriate for the region above the cut, and an outgoing
wave for
the region below. We choose the former condition, so in the end
we will obtain a retarded propagator. We denote the interior solution
by $\phi_1$, as in section 2.

Outside the shell the solution to equation (\ref{radeq})  cannot be
expressed in terms of known functions (except for the BTZ black hole).
so we use a WKB approximation. Again we consider the region $r\gg r_+\gg R$.
To start with, we rewrite the equation (\ref{radeq}) in an alternative form.
First, we rescale the field as
\be
\phi (r,\omega,k) = \frac{1}{\sqrt{r^{d-1}f(r)}}
\chi (r,\omega,k)
\ee
where $f(r)=f_2(r)= 1 +r^2 -\mu/r^{d-2}$.
Then, the equation for the function $\chi$ can
be written in a form which resembles a Schr\"odinger
equation
\be
-\partial_{rr} \chi^{(\pm)} + V(r) \chi^{(\pm)} =0
\ee
with a potential
\bea
V(r) &=& \frac{(d-1)(d-3)}{4r^2} +\frac{\partial^2_{rr}f(r)}{2f(r)}
-\frac{1}{4}\left(\frac{\partial_rf(r)}{f(r)}\right)^2 +\frac{1}{2}
\frac{(d-1)}{r}
\frac{\partial_rf(r)}{f(r)} \nonumber \\
\mbox{} & & \ -\frac{\omega^2 -(\frac{k^2}{r^2} +m^2)f(r)}{f^2(r)}
\eea
with $f(r)$ as defined above.

Outside the shell, we want to find the normalizable and non-normalizable
solutions $\chi^{\pm}$, which satisfy the boundary conditions
\be
\chi^{(\pm)} \approx r^{\frac{1}{2}\mp\nu},\ \ \ (r\rightarrow\infty)
\ee
Now $r\gg r_+$, and the potential reduces to
\be
V(r) = \frac{\nu^2-1}{r^2}-\frac{\omega^2-k^2}{r^4} \ ,
\ee
where $\nu=\sqrt{d^2/4+m^2}$ and $k^2=l(l+d-2)$ (where $l$ is a nonnegative integer).
In this case the equation can be solved in terms of Bessel functions,
and we find the normalizable and non-normalizable modes to be
\be
\chi^{(\pm)}(r,\omega,k) \approx 2^{\pm\nu} \Gamma(1\pm\nu)
(\omega^2-k^2)^{\mp\frac{\nu}{2}} \sqrt{r}
J_{\pm\nu}\left(\frac{\sqrt{\omega^2-k^2}}{r}\right)
\ee
However, in the region $(d^2-1) r^2 /4 \ll\omega^2-k^2 $, namely
for large frequencies, better
approximate solutions are obtained using the WKB expression
\be
\chi =\frac{1}{\tilde{V}^{1/4}} \left(A \exp(i\int\sqrt{\tilde{V}(r)}dr) +
       B \exp(-i\int\sqrt{\tilde{V}(r)}dr)\right)
\ee
with
\be
\tilde{V}(r) =  \frac{\omega^2}{(r^d-r_+^d)^2} 
  -\frac{k^2r^{d-4}}{r^d-r^d_+} \ .
\ee
Matching the WKB functions with the Bessel functions, the following
expressions are obtained for the normalizable and nonnormalizable modes:
\bea
\chi^{(\pm)}(r,\omega,k) &\approx& \sqrt{\frac{2}{\pi}}
\frac{1}{(\tilde{V}(r))^{1/4}}
2^{\pm\nu} \Gamma(1\pm\nu) (\omega^2-k^2)^{\mp\frac{\nu}{2}}\times
\nonumber \\
&& \times
\cos\left[ \frac{\sqrt{\omega^2-k^2}}{r} \left( 1+ 
\frac{r^d_+}{r^d} \frac{2\omega^2 -k^2}{2(\omega^2-k^2)} \right)
\mp\frac{\pi}{2}\nu-\frac{\pi}{4}\right]
\label{outside}
\eea
in the region $ (d-1)r^2/4 \ll\omega^2-k^2 $ and $r\gg r_+$.
These expressions will be used to find the propagator
using (\ref{match}).

Thus, for large frequencies, we can use the approximate solutions
(\ref{outside}) outside the shell. In the large frequency limit,
the Poincare mode $\phi_1$, discussed above, reduces to the form (we
can focus on $k=0$)
\be
\phi_1 \sim r^{\frac{1-d}{2}} e^{{\rm i}\frac{\omega}{r}}
\ee
up to some constants which cancel in the propagator.
Inserting the above expressions to the formula (\ref{match}),
we obtain the propagator (at $k=0$)
\be
G(\omega,k=0) \approx 2^{-2\nu} \frac{\Gamma(1+\nu)}{\Gamma(1-\nu)}
\omega^{2\nu}
\frac{\alpha e^{-{\rm i}\xi^{(-)}}e^{-i\pi\frac{\nu}{2}}
    + \beta e^{{\rm i}\xi^{(-)}} e^{+i\pi\frac{\nu}{2}} }                       {\alpha e^{-{\rm i}\xi^{(+)}} e^{+i\pi\frac{\nu}{2}}
 + \beta e^{{\rm i}\xi^{(+)}} e^{-i\pi\frac{\nu}{2}} }
\ee
with
\bea
\xi^{(\mp)} &=& \frac{\omega}{r_s} \left( 1+ \frac{r^d_+}{(d+1)r^d_s}
  \right) + \frac{\pi}{4}\\
\alpha &=& \frac{(1-d)}{2} \frac{r^d_+}{r_s^d} -\frac{2i\omega}{r_s} \\
\beta &=& \frac{(1-d)}{2} \frac{r^d_+}{r_s^d} \ .
\eea
A useful check is that for $r_+=0$ the last fraction in the propagator is
a constant, so the propagator correctly reduces to
the expected form in pure AdS.
The propagator has poles when the denominator
vanishes. For $r_s\gg r_+$
it follows that there are poles at the values of $\omega$ which satisfy
\be
\frac{\omega}{r_s} = \frac{3}{4} \pi +\frac{\nu}{2} -\frac{i}{2}
 \ln \left( 1 + \frac{4i\omega}{(d-1)r_s} \frac{r^d_s}{r^d_+} \right) \ .
\label{poles}
\ee
In the leading approximation, we drop the ``1'' inside the
brackets in the last term.
We obtain
\be
\omega _{n}\approx \frac{\pi
r_{s}}{R^2}\left( n+\frac{3}{4}+\frac{\nu }{2}\right) 
 -\frac{ir_{s}}{2R^2}
 \ln \left( \frac{4\pi n}{(d-1)}\frac{r_{s}^{d}}{r_{+}^{d}}\right) \ ,
\ee
where $n$ is a (large) positive integer. 
Notice that the imaginary part is negative since $r_s>r_+$.

Finally, we would like to emphasize we have also performed a calculation
using
the exact solutions in global and black hole coordinates, with regularity at
$r=0$
as the interior boundary condition. In that case, the resonances are found
by
examining the imaginary part of the propagator, the spectral function. They
are seen in the envelope that modulates the amplitudes of the spikes
associated
with the elementary excitations (the poles at the real values of $\omega$.
The
envelope has periods of hills and valleys, and by examining the position and
the width of the hill one can deduce the position of the complex poles.
In this case the calculations were performed numerically.
In this way one obtains a better approximation for the location of the
poles,
but the qualitative results remain the same. The approximations used in the
above give accurate enough results to reach the main conclusions in the case
of
large shells.

\end{document}